# Large anomalous Hall angle in a topological semimetal candidate TbPtBi

Jie Chen [1,2], Hang Li[2], Bei Ding[2], Hongwei Zhang[2], Enke Liu[2], Wenhong Wang[1,2*]

[1]Songshan Lake Materials Laboratory, Dongguan, Guangdong, China., Dongguan 100190, China

[2]State Key Laboratory for Magnetism, Beijing National Laboratory for Condensed Matter Physics and Institute of Physics, Chinese Academy of Sciences, Beijing 100190, China

*Corresponding author: wenhong.wang@iphy.ac.cn

**Abstract:**

The magnetotransport properties in antiferromagnetic half-Heusler single crystals of TbPtBi, a magnetic-field-induced topological semimetal with simple band structure, are investigated. We found that a nonmonotonic magnetic field dependence of the anomalous Hall resistivity in a high magnetic field (B>7T), which come from the change of band structure induced by the Zeeman-like splitting when applying the external magnetic field. The experiment results show that credible anomalous Hall resistivity and conductivity reach up to $0.6798 m\Omega cm$ and $125 \Omega^{-1} cm^{-1}$, respectively. A large AHA up to 33% is obtained in TbPtBi, which is comparable to typical ferromagnetic Weyl semimetal. The analysis of results show it should be attributed to topological band around $E_F$ and low carrier density.

Magnetotransport properties of magnetic materials receive considerable interests for their importance in basic science and for their potential in technological applications[1]. Anomalous Hall Effect (AHE), as an important topic in condensed matter, contains abundant physics for their many different mechanisms[2], such as intrinsic mechanism connected to Berry curvature in entangled Bloch electronics bands[3], extrinsic mechanism including skew scattering[4,5] and side jump scattering[6]. Recently, the rapid development of magnetic topological materials has greatly promoted the studies on AHE. Magnetic topological materials, such as $Co_3Sn_2S_2$[7,8], $Co_2MnGa$[9], $Co_2MnAl$[10], $Fe_3GeTe_2$[11], $PrAlGe$[12] and $Mn_3Sn$[13], show a large intrinsic AHE. All of these studies identify special band structures like Weyl node, opened nodal line in Weyl semimetal or nodal line semimetal host large Berry curvature. For their close relation between topological band structure and Berry curvature, the AHE is sensitive to the position of topological band relative to Fermi level $E_F$. However, it is not easy to realize this tunability of the band structure by the direct Zeeman splitting of external magnetic field for that the energy change is too small, about several meV[14]. For rare earth element R, the interactions between localized R spins through the RKKY (Ruderman-Kittel-Kasuya-Yosida) mechanism can produce an exchange field on the 5d orbital that make band show Zeeman-like splitting reaching to the order of 0.1eV[14]. When the topological band is close to $E_F$, the splitting will tune the band. Therefore, the $B$ dependence of tunable AHE is usually observed in rear-earth-based compounds，such as $EuTiO_3$[1] and half-Heusler compounds $GdPtBi$[15,16]. As a magnetic-field-induced Weyl semimetal[17,18], GdPtBi shows unconventional anomalous Hall effects[15,16].

On the other hand, as the sister compound of GdPtBi, antiferromagnetic (AFM) half-Heusler TbPtBi shows the large chiral-anomaly-induced negative magnetoresistance, the typical feature of Weyl semimetal[19]. Similar to GdPtBi, TbPtBi is magnetic-field-induced Weyl semimetal that should show a large AHE. Here, in this letter, we show an obvious abnormal behavior in Hall resistivity under high magnetic field, $B>7T$ which enable us to extract AHE. As a result, we obtained the anomalous Hall angle (AHA) as large as 33%, which is compare to the typical magnetic Weyl semimetal $Co_3Sn_2S_2$[7,20] and full Heusler compounds $Co_2MnGa/Al$[9,21].

We crystalized TbPtBi single crystal by Bi flux method[22,23] with a molar ratio Tb: Pt: Bi =1: 1: 20. Magnetotransport and magnetic properties were measured on the Steady High Magnetic Field Facilities in High Magnetic Field Laboratory and Quantum Design Physical Property Measurement System (PPMS). Chemical composition was examined by the scanning electron microscope with an energy-dispersive X-ray spectrometry (EDS). We defined the orientation of single crystal by X-ray diffraction. The DFT calculations were performed with WIEN2k code[24]. Perdew–Burke–Ernzerhof of Generalized Gradient Approximation (PBE-GGA) is used for the calculation of the exchange correlation potentials[25]. A large exchange parameter $U_{eff}$=0.6Ry was applied to rear-earth elements Tb, which would shift the 4f electrons far away from the Fermi level $E_F$.

As shown in Fig. 1(a), TbPtBi crystalizes in a MgAgAs-type structure with space group F-43m and lattice parameter is a=6.702Å. X-ray pattern in Fig. 1(b) identified the [001] orientation of the single crystal. The component of single crystal, prepared by Bi-flux method, is detected by EDS. Fig. 1(c) shows color maps of Tb, Pt, Bi. The elements are uniformly distributed on the surface. We got the nonstoichiometric TbPtBi single crystals with mole ratio of Tb: Pt: Bi=1.11: 1: 1. Fig. 1(d) presents the temperature (*T*) dependence of magnetization (*M*) under *B*=0.5T along [001] axis. The *MT* curve show an antiferromagnetic transition at 3.38K. The enlarged magnetic transition in low *T* is shown in the inset. The fitting of paramagnetic curve by Curie-Weiss law indicates a large effective magnetic moment $\mu_{eff}$=9.57$\mu_B$ which is close to theoretical value $\mu(Tb^{3+})$ =9.72$\mu_B$. The isothermal magnetization curves show a linear behavior in both antiferromagnetic (*T*<3.38K) and paramagnetic states. There is no metamagnetic transition in whole magnetic field range (*B*<9T). Therefore, magnetic structure as the origin of AHE can be excluded. Fig. 1(f) shows zero-field resistivity with current *I*//[100] for sample #1, #2 and #3. Semiconductor-to-metallic behavior indicated a gapless semiconductor or semimetal, which is coincident with semimetallic band structure in Fig. 2.

Indeed, a detailed calculation on the band structure and topological properties of TbPtBi under different magnetic states has been reported by Zhu *et al*[26]. The results show that TbPtBi is an AFM topological insulator, but Weyl semimetal state will emerge when the Tb magnetic moments are aligned along to [001]. Because *T*-range of AHE main distribute in paramagnetic state. Here, we illustrated its band structures and density of states (DOS) for paramagnetic (PM) and ferromagnetic (FM) states in Fig. 2 to show the change of band structure and DOS under the magnetic field. In PM state, because spin polarization is absence, the magnetic moment is fixed to zero. As shown in Fig. 2(a), band structure in PM state is similar to no magnetic half-Heusler LuPtBi[27], it shows an electron pocket and a hole pocket around Γ point. When Tb magnetic moment are aligned ferromagnetically, the bands around Γ point separate into four bands and the energy shift is about 0.09eV and the electron pocket and hole pocket still persist. For the nonstoichiometric component of our sample, the Fermi level $E_F$ may be shift. According to experiment results of transport, electron dominant the transport and obviously one carrier model indicate $E_F$ shift to a high energy level. As shown in Fig. 2(c), the DOS values are almost unchanged in the range from -50meV to 50meV. It implies that the carrier density may hold a constant value in this energy window. This conclusion provides us a possibility to exclude the influence of carrier density's change under external magnetic field.

Next, we performed Hall resistivity $\rho_{xy}$ and magnetoresistance $\rho_{xx}$ measurements on three TbPtBi samples with *B*//[001] and *I*//[100]. Fig. 3 presents the AHE and the progress of extracting $\rho_{xy}^A$ for sample #1. Similar to other half-Heusler compounds, sample #1 (see Fig. 3(a)) shows abnormal magnetoresistance curves in low magnetic field which can be attributed to the quantum coherence effect[22]. With increasing of external magnetic field, electron will decoherence and back to classical Drude model. For $\rho_{xx}$ of *T*=1.8K, the abnormal behavior disappears at 3T and shows a linear behavior in higher field. Interestingly, the linear $\rho_{xx}(B)$ curves in high field is overlapped for different *T*. Fig. 3(b) presents the Hall resistivity $\rho_{xy}$ under *B* up to 32T. In general, $\rho_{xy}$ of a magnetic material can be expressed as $\rho_{xy} = R_H B + \rho_{xy}^A$, where $R_H B$ represents normal Hall resistivity $\rho_{xy}^N$ due to Lorentz effect. In light orange area ($B<7T$), $\rho_{xy}$ show linear dependence on $B$ that means the normal Hall effect. With increasing $B$, there is an additional Hall signal which is supposed to be anomalous Hall resistivity $\rho_{xy}^A$. Obviously, there is a critical field $B_c$ for observing AHE and $B_c$ value is increased with increasing $T$. For Hall resistivity at 50K, the linear part persists up to about 15T. It indicates that higher $B$ was need to induce the additional Hall signal for high $T$. The origin of this critical field phenomenon may associate with the change of band structure.

To scale the contribution of $\rho_{xy}^A$, we fitted linear curve in light orange area for normal Hall signal $\rho_{xy}^N$ and extracted $\rho_{xy}^A$ by subtracting the fitting line. Fig. 3(c) takes Hall resistivity of 1.8K as an example to present this progress. The separation of $\rho_{xy}^A$ and $\rho_{xy}^N$ is clean and reliable. We estimate the peak of $\rho_{xy}^A(1.8K)$ is $0.6798 m\Omega cm$ which is about 3.7 times of $\rho_{xy}^A(2K)=0.18\ m\Omega cm$ of GdPtBi[16]. The inset is the fitting line for normal Hall resistivities at different $T$. The overlapped lines indicate the carrier concentration $n_e$ is almost unchanged in $T<50K$. The negative Hall factors $R_H$ imply that carrier of electron dominates the transport which is opposite to hole in previous data[26,28]. This differentiation can be attributed to the influence of nonstoichiometric component on the Fermi level E$_F$. The nonstoichiometric component may push the Fermi level a slightly higher than stoichiometric one. In this case, electron dominant the transport and only the conductance bands cross Fermi level. The obtained carrier concentration $n_e$ and mobility $\mu_e$ are shown in Table S1. The carrier concentration and mobility can reach $n_e = 6.41 \times 10^{18}\ cm^{-3}$, $\mu_e = 2037\ cm^{-2}V^{-1}s^{-1}$ at 1.8K. $n_e$ is same order to other half-Heusler RPtBi, but about half of previous result $n_h = 1.2 \times 10^{19}\ cm^{-3}$ in TbPtBi[28]. Fig. 3(d) shows the extracted $\rho_{xy}^A$ for different $T$ and $B$. There is big swell in the range of $B>7T$ and it persists to 32T. The inset shows the maximum values of $\rho_{xy}^A$ for corresponding $T$ and $\rho_{xy}^A$ decreases quickly with increasing $T$.

To further characterize AHE in TbPtBi, we employ two characteristic parameter,

anomalous Hall conductivity (AHC), $\sigma_{xy}^A$ and AHA. AHC is given as $\sigma_{xy}^A = -\rho_{xy}^A/(\rho_{xy}^2 + \rho_{xx}^2)$, which is absolute scale of AHE. Fig. 4(a) shows the dependence of $\sigma_{xy}^A$ on *T* and *B*. The orange area represents a large anomalous Hall conductance $\sigma_{xy}^A$. Compared to GdPtBi's result[16], the $\sigma_{xy}^A$ mainly distribute to a higher field range, 10T<*B*<22T and the maximum of $\sigma_{xy}^A$ reaches to $-125\Omega^{-1}cm^{-1}$. Fig. 4(b) shows the $\sigma_{xy}^A$ versus $\sigma_{xx}$ curve for *T*-range of 1.8-50K. We find that the relation of $\sigma_{xy}^A$ and $\sigma_{xx}$ can be divided in two parts. In the range of *T*>10K, $\sigma_{xy}^A$ shows a strongly dependence on $\sigma_{xx}$. However, $\sigma_{xy}^A$ is almost independent on $\sigma_{xx}$ below *T*=10K. This feature is consistent with intrinsic AHE mechanism. AHA is another parameter to scaling the AHE, defined by the ratio of $\sigma_{xy}^A/\sigma_{xx}$, which measures the relative contribution of the anomalous Hall current with respect to the longitudinal current[2]. TbPtBi, as a magnetic topological semimetal candidate, may feature a giant anomalous Hall angle for its special band structure. Fig. 4(c) shows the *T* and *B* dependence of AHA. Similar to anomalous Hall resistivity, AHA in the low field is close to zero. AHA shows a swell in *B*>7T. With decreasing *T*, AHA increases sharply from 5.1% at 50K to a maximum of 29% at 1.8K and 15T for sample #1.

In Fig. 5, we list the AHA in those famous compounds with large AHE. By comparison, we know that most of systems have a tiny AHA. Only a few of materials host a large value, like magnetic Weyl semimetal $Co_3Sn_2S_2$[7,20] and full Heusler compounds $Co_2MnGa/Al$[9,21]. Magnetic half-Heusler compounds RPtBi (like GdPtBi and TbPtBi) is another system with giant AHA. By analyzing, the common features of these system are topological band structure near Fermi level and semimetallic band. As we all known, topological band structure makes a great contribution to intrinsic AHC for their strong Berry curvature. The semimetallic band guarantee the low carrier density and small longitude conductivity. Both of these features are the requisite for getting giant AHA. For observation of chiral anomaly negative magnetoresistance, TbPtBi is regard as a magnetic-field-induced Weyl semimetal. Nevertheless, we need to point out that TbPtBi is not an intrinsic ferromagnetic compound, instead of a magnetic-field-induced ferromagnetic order. This feature makes the AHE exist in a high magnetic field, which is a disadvantage for applications. Therefore, reducing the critical field $B_c$ for AHE is the nest key challenge for realizing the applications.

This work was supported by the National Science Foundation of China (Nos.11974406 and 12074415). A portion of this work was performed on the Steady High Magnetic Field Facilities, High Magnetic Field Laboratory, Chinese Academy of Science.

**Supporting Information**

The data that support the finding of this study are available within this article.